\documentclass[sigconf]{acmart}
\usepackage{float}
\usepackage{booktabs}
\usepackage{xcolor}

\AtBeginDocument{%
  }

\copyrightyear{2026}
\acmYear{2026}
\setcopyright{cc}
\setcctype{by}
\acmConference[SIGGRAPH Posters '26]{Special Interest Group on Computer Graphics and Interactive Techniques Conference Posters}{July 19--23, 2026}{Los Angeles, CA, USA}
\acmBooktitle{Special Interest Group on Computer Graphics and Interactive Techniques Conference Posters (SIGGRAPH Posters '26), July 19--23, 2026, Los Angeles, CA, USA}
\acmDOI{10.1145/3799825.3818743}
\acmISBN{979-8-4007-2548-7/2026/07}
\begin{document}

\title{Context Aware AI Assistant and AR Interface for Lunar Extravehicular Activity (EVA) Procedural Guidance}

\settopmatter{authorsperrow=4}

\author{Rodrigo Gallardo}
\affiliation{
  \institution{MIT}
  \city{Cambridge}
  \state{MA}
  \country{USA}
}
\email{ragallar@mit.edu}

\author{Qilmeg Doudatcz}
\affiliation{
  \institution{MIT}
  \city{Cambridge}
  \state{MA}
  \country{USA}
}
\email{qlmg954@mit.edu}

\author{Ganit Goldstein}
\affiliation{
  \institution{MIT}
  \city{Cambridge}
  \state{MA}
  \country{USA}
}
\email{ganit@mit.edu}

\author{Ilkyaz Sarimehmetoglu}
\affiliation{
  \institution{MIT}
  \city{Cambridge}
  \state{MA}
  \country{USA}
}
\email{ilkyazs@mit.edu}

\author{Sergio Mutis}
\affiliation{
  \institution{MIT}
  \city{Cambridge}
  \state{MA}
  \country{USA}
}
\email{smutis@mit.edu}

\author{Alexander Htet Kyaw}
\affiliation{
  \institution{MIT}
  \city{Cambridge}
  \state{MA}
  \country{USA}
}
\email{alexkyaw@mit.edu}

\author{Anita Lin}
\affiliation{
  \institution{MIT}
  \city{Cambridge}
  \state{MA}
  \country{USA}
}
\email{anita899@mit.edu}

\author{Clara Emmerling}
\affiliation{
  \institution{MIT}
  \city{Cambridge}
  \state{MA}
  \country{USA}
}
\email{cemmer@mit.edu}

\author{Berfin Ataman}
\affiliation{
  \institution{MIT}
  \city{Cambridge}
  \state{MA}
  \country{USA}
}
\email{berfina@mit.edu}

\author{Skylar Tibbits}
\affiliation{
  \institution{MIT}
  \city{Cambridge}
  \state{MA}
  \country{USA}
}
\email{sjet@mit.edu}

\renewcommand{\shortauthors}{Gallardo et al.}
\authorsaddresses{}

\begin{abstract}
As human space exploration returns to the Moon, astronauts need rapid access to procedural information during extravehicular activities (EVAs), where attention is divided across navigation, repair tasks, tool handling, and environmental risk. The challenge is not the absence of information, but surfacing the right information at the right moment. We present GAIN-AI (Guided Assistant for Intelligent Navigation), a context-aware AI assistant and minimal heads-up interface for procedural guidance in simulated lunar EVA. The system operates in two layers. The first grounds a large language model with structured context: EVA procedure documents, live telemetry data, and error-handling protocols encoded as JSON. The second restructures that output into three compact units for AR display: Goal, Task, and Verification. Evaluated on 111 synthetic EVA scenarios, the system scores 10.0/10 on nominal conditions and 8.15/10 on single-fault scenarios, with performance degrading on multi-fault and boundary-threshold cases.
\end{abstract}




\begin{teaserfigure}
  \includegraphics[width=\textwidth]{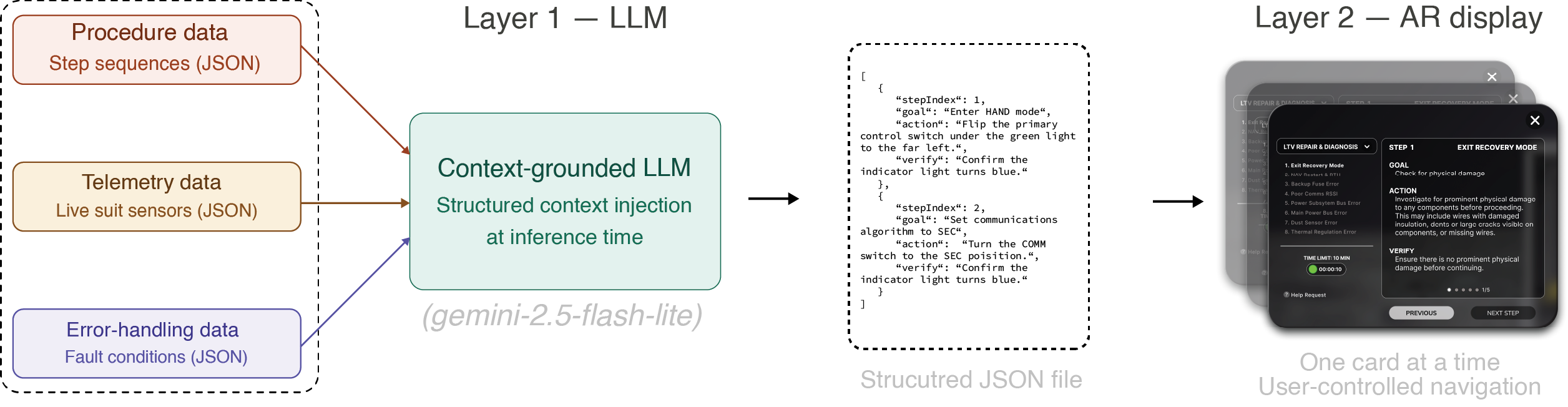}
  \caption{GAIN-AI deployed in a simulated lunar EVA scenario. The card-based heads-up display surfaces one actionable step at a time, reducing visual clutter while preserving procedural continuity.}
  \Description{GAIN-AI AR interface shown in a simulated lunar EVA environment.}
  \label{fig:teaser}
\end{teaserfigure}

\maketitle

%% INTRODUCTION
\section{Introduction}

Extravehicular activities are among the most cognitively demanding tasks astronauts perform, requiring sustained attention across mission objectives, environmental hazards, and multi-step procedures under physical constraint \cite{anderson_identifying_2025}. High cognitive workload is a documented safety risk \cite{anderson_characterizing_2026}, and existing procedures, though comprehensive and highly rehearsed, are difficult to use as real-time operational interfaces.

Prior work in AR-assisted EVA has demonstrated the value of heads-up display for biometrics, navigation, and communication \cite{thomas_augmented_2020, zhuang_2024_2025}, and recent research shows that LLM-AR integration can meaningfully reduce cognitive load during time-sensitive operations \cite{xu_integrating_2025}. Standard LLMs that lack domain-specific grounding often misinterpret user intent \cite{gallardo_scene-aware_2025, kyaw2025text}. This can be an issue in multi-fault situations or near boundary-threshold conditions that define real operational environments where the model’s reasoning diverges from the system’s actual state. Therefore, there is a need for LLMs to be connected to structured context, including mission procedures, live telemetry, and formalized error-handling knowledge, during runtime~\cite{gao_retrieval-augmented_2024}.

GAIN-AI addresses this gap with two layers: a general-purpose LLM grounded with structured JSON context, and a Goal-Task-Verification AR display that restructures model output into discrete, actionable cards. Our focus is not on replacing official procedures, but on reshaping them into a form that supports action under constraint.

%% SYSTEM OVERVIEW
\section{System Overview}

\begin{figure}[h!]
    \centering
    \includegraphics[width=1\linewidth]{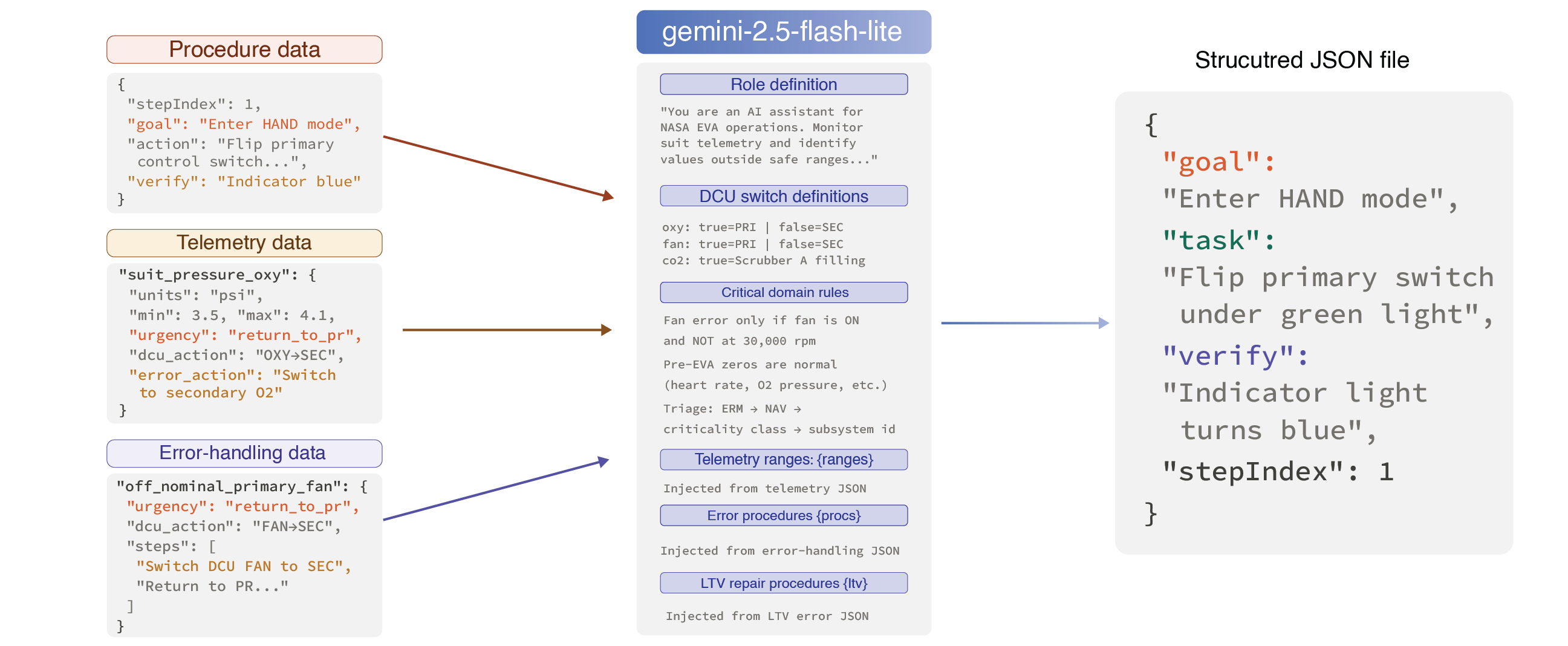}
    \caption{GAIN-AI system architecture. Procedure documents are parsed by the AI layer into Goal-Task-Verification units, rendered progressively on the AR heads-up display.}
    \Description{Diagram showing procedure, telemetry, and error-handling data entering the context-grounded language model, whose output is transformed into Goal, Task, and Verification cards for the AR display.}
    \label{fig:modules}
\end{figure}

\textbf{Layer 1: Context-grounded LLM.} A general-purpose LLM is provided with three structured JSON context sources at inference time: (1) \textit{procedure data}, encoding the full step sequence for a given EVA task; (2) \textit{telemetry data}, providing live suit and environmental sensor readings; and (3) \textit{error-handling data}, specifying fault conditions, severity thresholds, and corrective actions. We generated 111 synthetic EVA scenarios spanning nominal, single-fault, multi-fault, vehicle error, and boundary-threshold conditions, scoring model responses against ground-truth outcomes across four dimensions: error detection, correct action, urgency calibration, and triage ordering.

\textbf{Layer 2: Structured AR delivery.} Grounded output is restructured into discrete units, each containing a \textbf{Goal} (the objective of the current step), a \textbf{Task} (the specific action to perform), and a \textbf{Verification} condition (observable evidence of completion) \cite{zhao_guided_2025}. These are rendered as sequential cards on a Microsoft HoloLens~2 heads-up display, one at a time, with explicit user-controlled navigation, preserving operator agency while reducing the cost of parsing a full procedure document (Fig.~\ref{fig:modules}).

%% INTERFACE DESIGN
\section{Interface Design}

\begin{figure}[h!]
    \centering
    \includegraphics[width=1\linewidth]{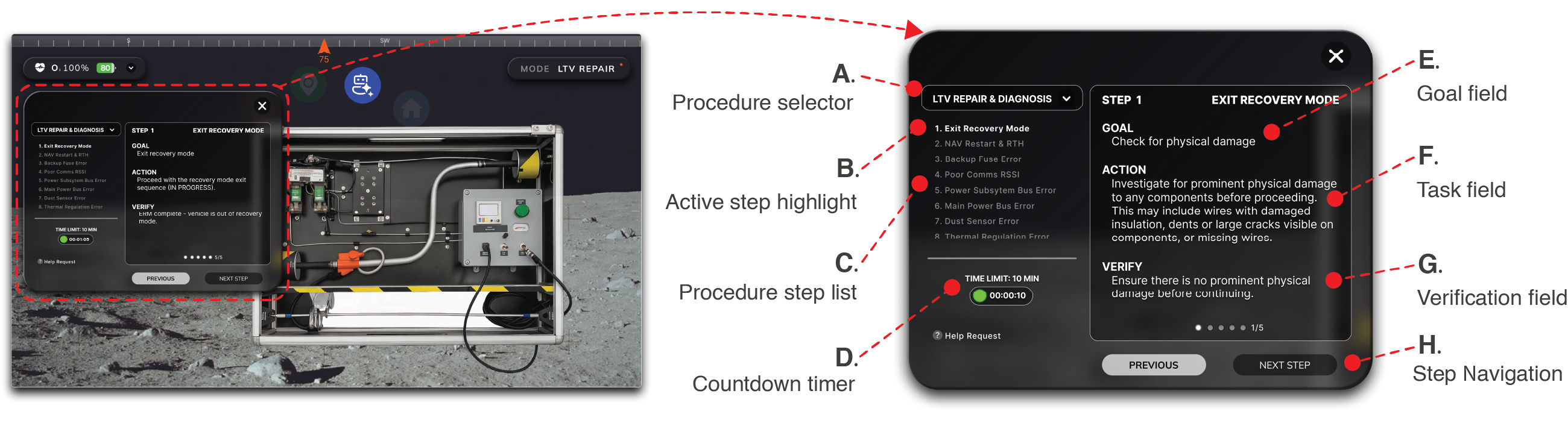}
    \caption{The GAIN-AI card interface. Each card displays a Goal, Task, and Verification field, prioritizing the current action while keeping the verification condition immediately accessible.}
    \Description{Three-part AR instruction card with labeled Goal, Task, and Verification fields.}
    \label{fig:ui}
\end{figure}

The interface was designed for conditions hostile to reading: a pressurized helmet with limited field of view, gloves that preclude fine motor interaction, and physical exertion competing for attention. Denser HUD displays produce measurable degradation in far-field situational awareness \cite{lee_visual_2024}, and providing more data does not monotonically improve outcomes; surfacing additional information can itself raise cognitive load~\cite{gupta_insights_2025}. Every design decision follows from this: legible at a glance, navigable with minimal input, with the relevant information always immediately visible.

Each card contains exactly three labeled fields (Fig.~\ref{fig:ui}). The \textbf{Goal} states the high-level objective in one sentence. The \textbf{Task} states the specific physical action in imperative form, limited to a single instruction. The \textbf{Verification} field states the observable condition confirming completion, a sensor reading, physical state, or system indicator. Elevating Verification to a primary element (rather than supplemental text) is deliberate: missed verifications are a common source of procedural error under time pressure, and pairing every Task with an explicit Verification cue reduces ambiguity at step transitions without requiring cross-reference to a longer document. The layout uses high-contrast typography with a clear label-value hierarchy and a palette designed for legibility across both the bright lunar surface and the lower-light airlock interior.

%% EVALUATION
\section{Evaluation}

We tested \textit{gemini-2.5-flash-lite}, selected for its multimodal capabilities and latency profile, with full structured context injection across 111 synthetic EVA scenarios in five categories: nominal, single-fault, multi-fault, LTV vehicle errors, and boundary-threshold edge cases. Each scenario was scored on error detection (max 3), correct action (max 3), urgency calibration (max 2), and triage ordering (max 2), for a maximum of 10 points~\cite{chang_survey_2023}.

\begin{table}[h!]
\centering
\footnotesize
\caption{LLM performance with structured context injection across EVA scenario categories. Per-dimension maxima: Detect.\ 3, Action 3, Urgency 2, Triage 2; Total 10.}
\label{tab:eval_results}
\begin{tabular*}{\columnwidth}{@{\extracolsep{\fill}}lcccccc@{}}
\toprule
\textbf{Category} & \textbf{N} & \textbf{Det.} & \textbf{Act.} & \textbf{Urg.} & \textbf{Tri.} & \textbf{Total} \\
\midrule
Nominal       & 12 & 3.00 & 3.00 & 2.00 & 2.00 & 10.00 \\
Single error  & 38 & 2.95 & 1.97 & 1.97 & 2.00 & 8.15 \\
Multi-error   & 30 & 2.73 & 1.83 & 2.00 & 1.90 & 7.73 \\
LTV errors    & 11 & 1.73 & 0.45 & 0.91 & 0.73 & 3.96 \\
Edge/bound.   & 20 & 1.90 & 1.35 & 1.30 & 1.95 & 5.24 \\
\midrule
\textbf{Overall} & \textbf{111} & \textbf{2.52} & \textbf{1.74} & \textbf{1.66} & \textbf{1.78} & \textbf{7.18} \\
\bottomrule
\end{tabular*}
\end{table}

Performance is strong on nominal (10.0/10) and single-fault (8.15/10) scenarios and degrades with complexity. Multi-fault drops to 7.73/10 from failures in action sequencing; LTV errors (3.96/10) and boundary cases (5.24/10) expose the limits of general JSON encoding: the 13-step NAV restart sequence demands tracking exact switch states, indicator colors, and reset timings in strict order, while boundary cases hinge on precise sensor thresholds (e.g., tank pressures above 3000\,psi or below 10\,psi) that the model rounds or misorders. Error detection remains stable across categories (avg. 2.52/3) while correct action drops sharply (avg. 1.74/3), indicating the model identifies anomalies but lacks the grounding to prescribe precise responses, the direct target of context injection. The evaluation framework and scenarios are released publicly.\footnote{\url{https://rodrigoagallardo008.github.io/EVA-LLM-Evaluator/}}

%% LIMITATIONS AND FUTURE WORK
\section{Limitations and Future Work}

\begin{figure}[h!]
    \centering
    \includegraphics[width=\linewidth]{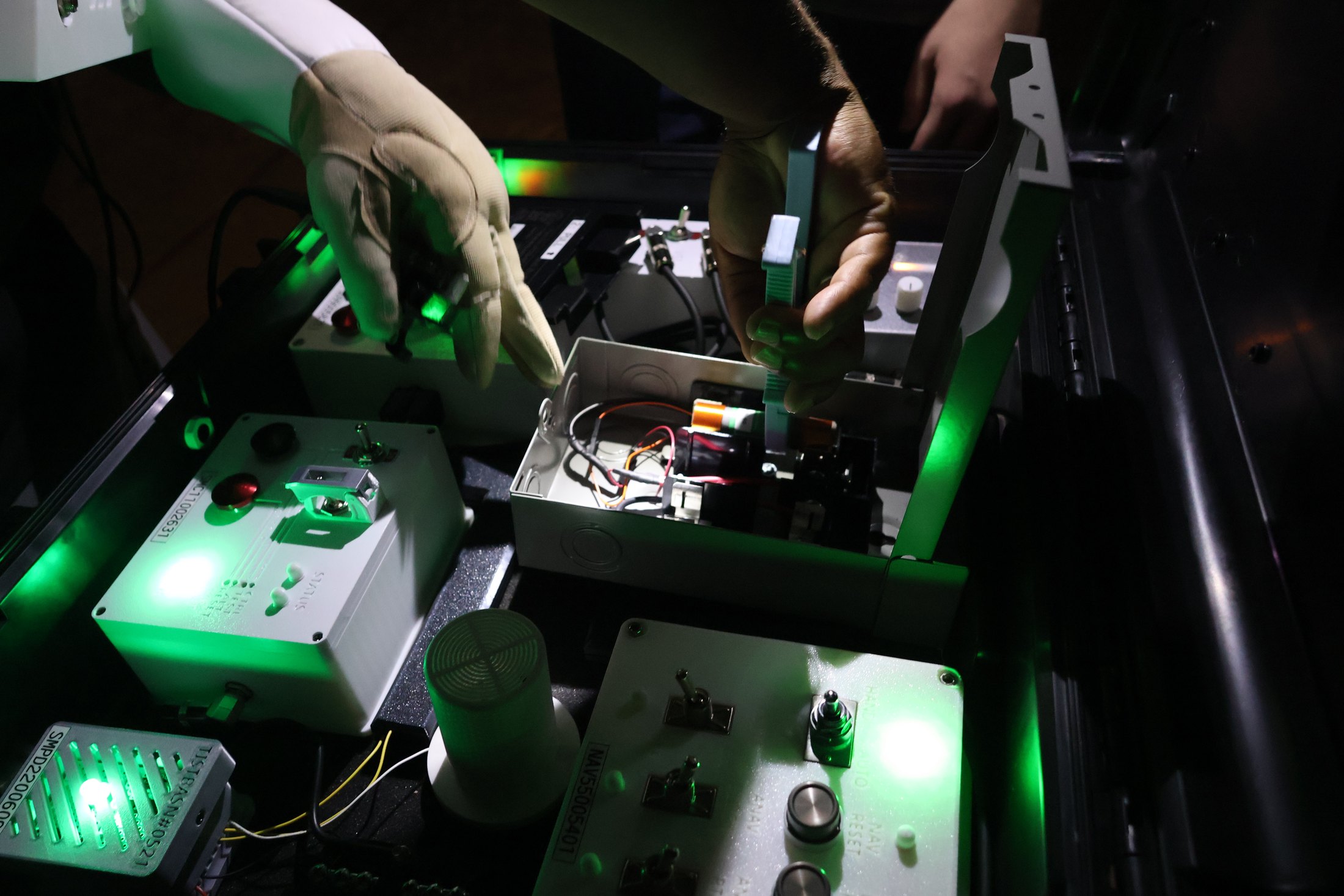}
    \caption{GAIN-AI deployed during a human-factors study at NASA Johnson Space Center, May 2026. An operator runs procedural guidance on a HoloLens~2 while performing EVA repair tasks; full results are forthcoming.}
    \Description{An operator wearing a HoloLens 2 and EVA chest control unit performs repair tasks at a test site while team members observe.}
    \label{fig:jsc}
\end{figure}

This work presents a design prototype and early evaluation framework, implemented as one subsystem within a larger EVA system that also includes a camera-free gesture recognition glove. The complete system was tested at Johnson Space Center in May 2026 as a human factors engineering study (Fig.~\ref{fig:jsc}); full results are forthcoming. The results here cover the LLM logic and UI design separately, so a full picture of system performance awaits analysis of the integrated study. The synthetic scenarios do not yet measure effects on real user performance or cognitive workload. Future work will add a no-injection baseline to complete the comparison, integrate live telemetry for real-time updates, and conduct user studies comparing the GTV display against conventional checklists, with multimodal hands-free input as a longer-term target.

\begin{acks}
The authors thank the NASA SUITS team for their support throughout this project, and Jamarius Reid from Axiom Space for his mentorship and guidance.
\end{acks}

\bibliographystyle{ACM-Reference-Format}
\bibliography{references}

\end{document}